\documentclass[12pt]{article}
\usepackage{graphicx}
\pdfoutput=1

\newcommand\pubnumber{UCHEP--15--03}
\newcommand\pubdate{\today}

\def\UCphysics{Physics Department\\
University of Cincinnati, Cincinnati, Ohio 44221 USA}

\def\Title#1{\begin{center} {\Large #1 } \end{center}}
\def\Author#1{\begin{center}{ \sc #1} \end{center}}
\def\Address#1{\begin{center}{ \it #1} \end{center}}

\newcommand\pubblock{\rightline{\begin{tabular}{l} \pubnumber\\
         \pubdate  \end{tabular}}}
\newenvironment{Abstract}{\begin{quotation}  }{\end{quotation}}
\newenvironment{Presented}{\begin{quotation} \begin{center} 
             PRESENTED AT\end{center}\bigskip 
      \begin{center}\begin{large}}{\end{large}\end{center} \end{quotation}}
\def\Acknowledgements{\bigskip  \bigskip \begin{center} \begin{large}
             \bf ACKNOWLEDGEMENTS \end{large}\end{center}}

\def\cp{$CP$\/}
\def\cpv{$CPV$\/}

\def\ycp{$y^{}_{CP}$\/}
\def\agamma{$A^{}_{\Gamma}$\/}

\def\ra{\!\rightarrow\!}

\def\dbar{\overline{D}{}^{\,0}}

\def\pt{p^{}_T\/}

\def\dklnu{$D^0\ra K^+\ell^-\nu$}
\def\dkpi{$D^0\ra K^+\pi^-$}
\def\dkk{$D^0\ra K^+K^-$}
\def\dpipi{$D^0\ra\pi^+\pi^-$}

\def\dkspp{$D^0\ra K^0_S\,\pi^+\pi^-$}
\def\dkskk{$D^0\ra K^0_S\,K^+ K^-$}

\def\meve{~MeV}

\def\mevm{~MeV/$c^2$}

\def\gevm{~GeV/$c^2$}

\def\beq{\begin{equation}}
\def\eeq#1{\label{#1}\end{equation}}
\def\eeqn{\end{equation}}

\def\beqa{\begin{eqnarray}}
\def\eeqa#1{\label{#1}\end{eqnarray}}
\def\eeqan{\end{eqnarray}}

\let\bar=\overbar

\def\Dslash{\not{\hbox{\kern-4pt $D$}}}
\def\dslash{\not{\hbox{\kern-2pt $\del$}}}

\def\msb{{\bar{\ssstyle M \kern -1pt S}}}

\begin{document}
\begin{titlepage}
\pubblock

\vfill
\Title{CHARM 2015 Experimental Summary: Step-by-Step Towards New Physics}
\vfill
\Author{A. J. Schwartz}
\Address{\UCphysics}
\vfill
\begin{Abstract}
The experimental program of the 
Seventh International Workshop on Charm Physics (CHARM 2015)
is summarized. Highlights of the workshop include results
from heavy flavor production, quarkonium and exotic states,
hadronic decays and Dalitz analyses,
semileptonic and leptonic decays,
rare and radiative decays,
charm mixing, and 
\cp\ and $T$ violation.
\end{Abstract}
\vfill
\begin{Presented}
The 7th International Workshop on Charm Physics (CHARM 2015)\\
Detroit, MI, 18-22 May, 2015
\end{Presented}
\vfill
\end{titlepage}
\def\thefootnote{\fnsymbol{footnote}}
\setcounter{footnote}{0}
\setcounter{page}{2}

\section{Introduction}

This year's workshop featured almost fifty experimental
talks covering a wide variety of charm physics results. The
presentations can be categorized as follows:
heavy flavor production;
quarkonium and exotic states;
hadronic decays and Dalitz analyses;
semileptonic decays; leptonic decays;
rare and radiative decays;
\cp\ violation and mixing;
beyond-the-Standard-Model searches;
charm baryons, $\tau$ leptons, and other miscellaneous decays.
Results were presented from Belle, BaBar, CLEOc, BESIII, CDF,
D0, ATLAS, CMS, LHCb, STAR, PHENIX, and ALICE experiments. 
In this review I discuss some of the highlights from among
these talks. Space constraints allow for only a cursory discussion,
and for more details the reader is referred to the original
presentation and writeup.

\section{\boldmath Production}
\label{sec:production}

\begin{figure}[htb]
\vskip-0.8in
\centering
\includegraphics[width=116mm,angle=-90]{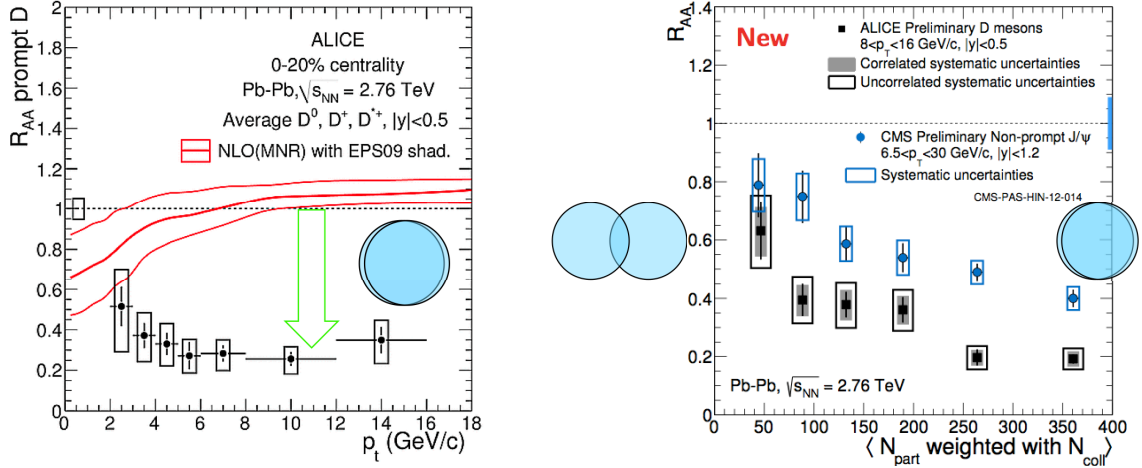}
\vskip-1.0in
\caption{
Left: $R^D_{AA}$ measured by ALICE in Pb-Pb collisions
using $D^0\ra K^-\pi^+$ decays.
Right: $R^{}_{AA}$ plotted vs.\ $N^{}_{\rm part}$, where
higher (lower) values correspond to central (peripheral)
collisions. Included in this plot is $R^B_{AA}$ measured by CMS
in Pb-Pb collisions using non-prompt $J/\psi$ decays~\cite{hi_cms}.
The data shows that $R^B_{AA}>R^D_{AA}$, as expected. }
\label{fig:suppression}
\end{figure}

Numerous results were presented on $D$ meson production in proton-nucleus
and nucleus-nucleus collisions (Dainese). The temperatures of such collisions
correspond to the environment of a quark-gluon plasma (QGP), and partons
produced in the collision interact with the QGP when escaping,
resulting in an energy loss. The parton energy loss is expected to
decrease as the parton mass increases. The parameter quantifying 
parton interaction with the QGP is the ``nuclear modification factor:''
\begin{eqnarray}
R_{AA}(\pt) & \equiv &  \frac{1}{N^{}_{\rm coll}} 
\frac{(dN^{}_{AA}/dp^{}_T)}{(dN^{}_{pp}/dp^{}_T)}\,,
\end{eqnarray}
where $N~{}_{\rm coll}$ is the number of nucleons participating
in the collision.
The greater the interaction with the QGP and subsequent energy loss,
the lower $R_{AA}(\pt)$; this is referred to as ``suppression.'' 
The amount of suppression is found to increase with $p^{}_T$; typical
behavior is shown in Fig.~\ref{fig:suppression}(left), which 
plots Pb-Pb data taken by the ALICE experiment. 
Suppression also depends upon
the nuclei colliding: using $\pi^0$ production and also electrons
from heavy flavor decays, PHENIX shows that in Au-Au collisions
suppression is significant, but in $d$-Au collisions it is not. 
ALICE confirms this trend for $D^0$ production by reconstructing
$D^0\ra K^-\pi^+$ decays~\cite{charge-conjugates}: in Pb-Pb
collisions $D^0$ suppression is significant, while in $p$-Pb
collisions it is not. 

Finally, suppression also depends on the ``centrality'' of a collision,
i.e., the amount of overlap of the colliding hadronic systems. The
centrality of a collision is inferred from the multiplicity of
secondary particles produced: a high multiplicity of secondaries
indicates large hadronic overlap. Data from ATLAS shows that
suppression is largest for central collisions and smallest
for peripheral collisions, for several ranges of $\pt$ studied. 

CMS reconstructs $J/\psi$ decays in Pb-Pb collisions and, by
requiring that the $J/\psi$ have large impact parameter, identifies
these $J/\psi$'s as originating from $B$ decays~\cite{hi_cms}.
Comparing $R_{AA}$ for this $B$ sample with $R_{AA}$ measured
by ALICE for $D$ decays shows that $R^B_{AA}>R^D_{AA}$ 
[Fig.~\ref{fig:suppression}(right)], as expected because
$m^{}_b > m^{}_c$. This is an important confirmation of
this relationship.

\section{\boldmath X/Y/Z Quarkonia}

Results for $X/Y/Z$ states were presented by BESIII, Belle,
and ATLAS. A subset of these results are summarized below.

\subsection{BESIII}

BESIII (Kornicer, Lyu, Prasad) selects
$e^+e^-\ra (J/\psi,h^{}_c) \pi^+\pi^-$ and $(J/\psi, h^{}_c)\pi^0\pi^0$
decays, where $J/\psi\ra\ell^+\ell^-$ and $h^{}_c(1P)\ra\eta^{}_c\gamma$, 
and plots the invariant mass of the $J/\psi\,\pi$ and
$h^{}_c \pi$ combinations. In the $J/\psi$ sample, prominent
peaks are observed for $Z^{}_c(3900)^+$ and $Z^{}_c(3900)^0$ 
states; in the $h^{}_c$ sample, peaks are observed for
$Z^{}_c(4020)^+$ and $Z^{}_c(4020)^0$ states. The fitted masses
with systematic errors are listed in Table~\ref{tab:besIII_zmass}.

\begin{table}[hbt]
\renewcommand{\arraystretch}{1.1}
\begin{center}
\begin{tabular}{l|c}  
State &  Fitted mass \\
\hline
$M^{}_{Z^{}_c(3900)^+}$ & $3899.0\pm 3.6\pm 4.9$ \\
$M^{}_{Z^{}_c(3900)^0}$ & $3894.8\pm 2.3\pm 2.7$ \\
$M^{}_{Z^{}_c(4020)^+}$ & $4022.9\pm 0.8\pm 2.7$ \\
$M^{}_{Z^{}_c(4020)^0}$ & $4023.9\pm 2.2\pm 3.8$ \\
\hline
\end{tabular}
\caption{States observed by BESIII in
$e^+e^-\ra (J/\psi,h^{}_c) \pi^+\pi^-$ and $(J/\psi, h^{}_c)\pi^0\pi^0$
reactions, and the fitted masses with systematic errors.}
\label{tab:besIII_zmass}
\end{center}
\end{table}

BESIII also selects 
$e^+e^-\ra \pi^-(D\overline{D}^*)^+$ events.
Plotting the invariant mass of the 
$(D\overline{D}^*)^+$ pair shows an
enhancement just above threshold, which
may be an isospin partner of the $X(3872)$;
the fitted mass is $3884.3\pm 1.2\pm 1.5$.
Selecting $e^+e^-\ra J/\psi\,\pi^+\pi^-\gamma$
events and plotting the $J/\psi\,\pi^+\pi^-$ three-body
mass shows a sharp peak at the $X(3872)$; the fitted
mass is $3871.9\pm 0.7\pm 0.2$. BESIII measures
the cross section for this reaction at three
different center-of-mass energies and observes
a rise in the cross section that is consistent 
with a Breit-Wigner lineshape from the $Y(4260)$;
this is suggestive of $Y(4260)\ra X(3872)\gamma$
decays. The measured rate would correspond to
a ratio
${\cal B}(Y^{}_{4260}\ra X^{}_{3872}\gamma)/
{\cal B}(Y^{}_{4260}\ra J/\psi\,\pi^+\pi^-)\approx 0.10$,
which, if true, implies that the $X(3872)$ and
$Y(4260)$ are related.

Finally, BESIII reconstructs
$e^+e^-\ra \chi^{}_c \pi^+\pi^-\gamma$ events,
where $\chi^{}_c\ra J/\psi\,\gamma$.
Plotting the $\pi^+\pi^-$ recoil mass 
exhibits a new $X(3823)$ state 
[Fig.~\ref{fig:xyz}(left)]. The
significance is $6.7\sigma$, and
the fitted mass and width are 
$M=3821.7\pm 1.3 \pm 0.7$ and
$\Gamma < 16$~MeV at 90\% C.L. These values
are consistent with those expected for
the $\psi(1^3D_2)$ state, and BESIII
may be observing
$e^+e^-\ra\psi(1^3D_2)\pi^+\pi^-,\, 
\psi(1^3D_2)\ra \chi^{}_c \gamma$
reactions.

\subsection{Belle}

Belle (Wang, Bhardwaj) showed results for $X/Y/Z$ states produced
in $B$ decays. The decays $B^\pm\ra K^\pm \eta^{}_c h$ are reconstructed,
where $h = \omega,\,\eta,\,\pi^0,\,\pi^+\pi^-$, and the $\eta^{}_c h$
mass spectrum is studied for evidence
of $X(3730),\,X(3872),\,X(3915),\,X(4014),\,Z(3900)$ and $Z(4020)$
states. No evidence is seen for any of these states, and upper
limits are set for the 
branching fractions $B^\pm\ra K^\pm (X,Z),\,(X,Z)\ra\eta^{}_c h$.
These limits lie in the range $(1.2-6.9)\times 10^{-5}$.

\begin{figure}[htb]
\centering
\hbox{
\hskip-0.10in
\includegraphics[width=56mm,angle=-90]{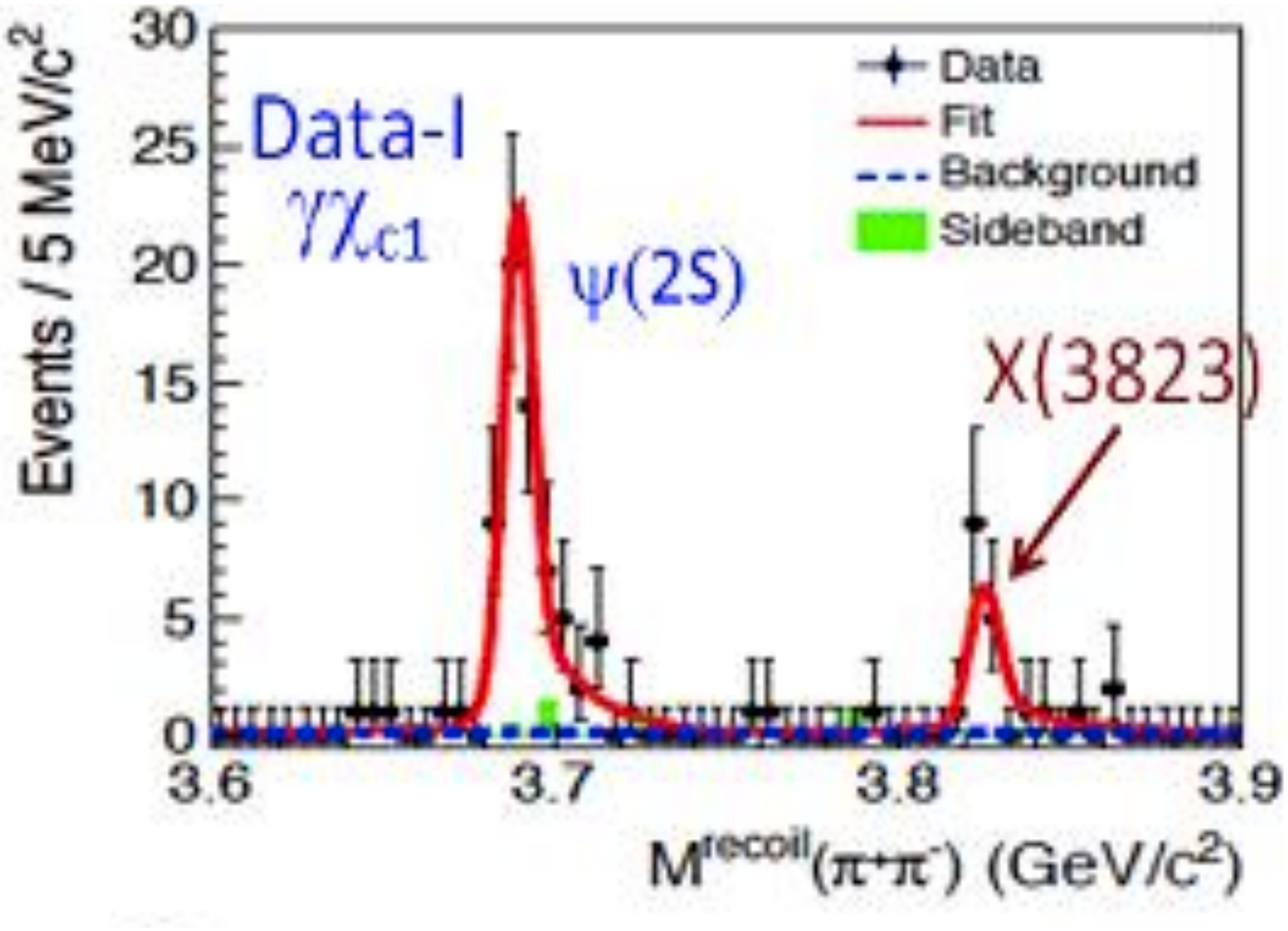}
\hskip0.30in
\includegraphics[width=52mm,angle=-90]{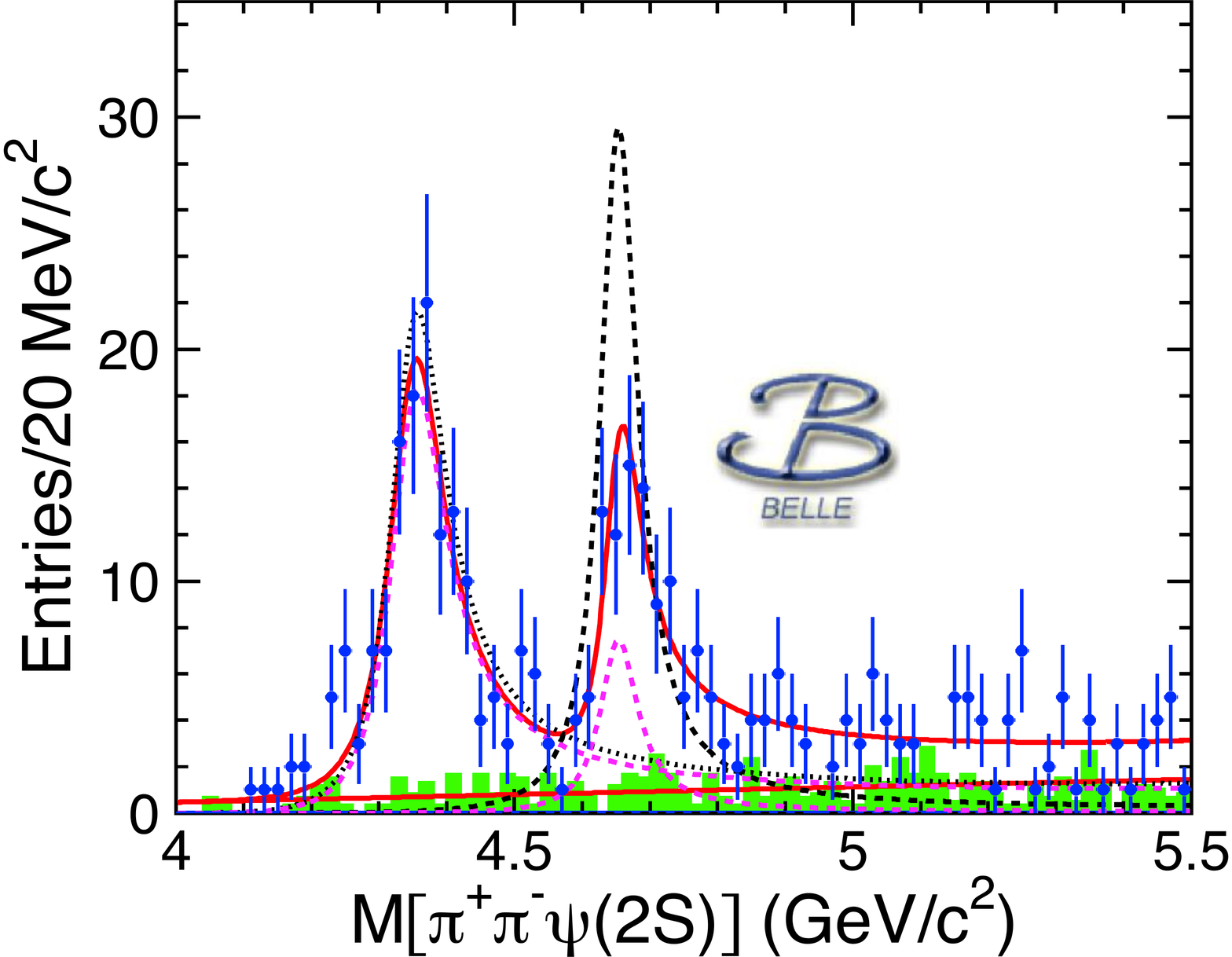}
}
\caption{Left: BESIII measurement of the $\pi^+\pi^-$ recoil
mass in $e^+e^-\ra \pi^+\pi^-\chi^{}_c\gamma$ reactions.
The rightmost peak indicates $X(3823)\ra\chi^{}_c\gamma$
decays.
Right: Belle measurement of the $\psi(2S)\pi^+\pi^-$
mass spectrum in $B^+\ra \psi(2S)\pi^+\pi^- K^+$ decays,
where $\psi(2S)\ra (J/\psi\,\pi^+\pi^-,\,\mu^+\mu^-)$. The
peaks indicate $Y(4360)\ra\psi(2S)\pi^+\pi^-$ and
$Y(4660)\ra \psi(2S)\pi^+\pi^-$ decays. The results
of fitting to Breit-Wigner amplitudes are superimposed.
}
\label{fig:xyz}
\end{figure}

Belle also presented results for $Y(4360)$ and $Y(4660)$ production
and decay.
First, $B^+\ra \psi(2S)\pi^+\pi^-K^+$ decays are reconstructed
and the $\psi(2S)\pi^+\pi^-$ mass distribution is plotted,
where $\psi(2S)\ra J/\psi\,\pi^+\pi^-$ or $\mu^+\mu^-$.
Both mass distributions show prominent peaks for
the $Y(4360)$ and $Y(4660)$ states; the combined 
distribution is shown in Fig.~\ref{fig:xyz}(right). 
Fitting the peaks to Breit-Wigner amplitudes yields
the parameters listed in Table~\ref{tab:belle_psiprpp}.
In a second analysis the $Y(4360)$ is produced directly
via $e^+e^-\ra \psi(2S)\pi^+\pi^-$~\cite{belle_wang}.
The final state is reconstructed as a function of center-of-mass
energy ($\sqrt{s}$), and peaks are seen at $\sqrt{s}\approx 4350$\meve\ 
and 4650\meve. The $Y(4360)$ peak corresponds to a cross section
of $\sim$\,75~pb, which is surprisingly close to the cross sections
measured previously by Belle for $e^+e^-\ra Y(4260)\ra J/\psi\,\pi^+\pi^-$
and $e^+e^-\ra \psi(4040)\ra J/\psi\,\eta$
reactions~\cite{belle_4260,belle_4040}.

\begin{table}[hbt]
\renewcommand{\arraystretch}{1.1}
\begin{center}
\begin{tabular}{l|cc}  
State &  Fitted mass (\mevm) & Fitted width (\meve) \\
\hline
$Y(4360)$ & $4347\,\pm 6\,\pm 3$ & $103\,\pm 9\,\pm 5$ \\
$Y(4660)$ & $4652\,\pm 10\,\pm 11$ & $68\,\pm 11\,\pm 5$ \\
\hline
\end{tabular}
\caption{Parameters measured by Belle by fitting
the $\psi(2S)\pi^+\pi^-$ invariant mass spectrum in
$B^\pm\ra \psi(2S)\pi^+\pi^-K^\pm$ decays.}
\label{tab:belle_psiprpp}
\end{center}
\end{table}

Finally, Belle searched for $X(3872)$ production via
$B^0\ra X(3872) K^+\pi^-$ and
$B^+\ra X(3872) K^0_S\,\pi^+$, where $X(3872)\ra J/\psi\,\pi^+\pi^-$.
In both cases clear peaks are observed above background; the signal
significances are $7.0\sigma$ and $3.7\sigma$, respectively. The
number of signal candidates for the more copious neutral mode
is $151\pm 21$, and the product of branching fractions is
${\cal B}(B^0\ra X(3872) K^+\pi^-)\times {\cal B}(X(3872)\ra J/\psi\,\pi^+\pi^-)
= (7.9\,\pm 1.3\,\pm 0.4)\times 10^{-6}$.
The $K^+\pi^-$ mass spectrum is subsequently fitted
to identify $K^*(890)$ production, and the resonant
fraction is measured to be
\begin{eqnarray}
\frac{{\cal B}(B^0\ra X(3872)K^{*0})\times {\cal B}(K^{*0}\ra K^+\pi^-)}
{{\cal}{\cal B}(B^0\ra X(3872)K^+\pi^-)} & = & 0.34\,\pm 0.09\,\pm 0.02\,.
\end{eqnarray}

\section{\boldmath Hadronic Decays}

Results for hadronic decays, including several Dalitz analyses,
were presented by BESIII, BaBar, and LHCb, as summarized below.

\subsection{BESIII}

BESIII (Wiedenkaff, Muramatsu) reported measurements of
singly Cabibbo-suppressed decays of $D$ mesons:
$D^+\ra\omega\pi^+$, 
$D^0\ra\omega\pi^0$, 
$D^+\ra\eta\pi^+$, and
$D^0\ra\eta\pi^0$. The results are listed in Table~\ref{tab:hadronic_besiii}.
For the final states with $\omega$, these results provide the first
evidence for these decays.

\begin{table}[hbt]
\renewcommand{\arraystretch}{1.1}
\begin{center}
\begin{tabular}{l|cc}  
Mode &  Branching fraction & PDG value \\
\hline
$D^+\ra\omega\pi^+$  & $(2.74\,\pm 0.58\,\pm 0.17)\times 10^{-4}$ & 
     $<3.4\times 10^{-4}$ (90\% C.L.) \\
$D^0\ra\omega\pi^0$  & $(1.05\,\pm 0.41\,\,\pm 0.09)\times 10^{-4}$ & 
     $<2.6\times 10^{-4}$ (90\% C.L.) \\
$D^+\ra\eta\pi^+$  & $(3.13\,\pm 0.22\,\pm 0.19)\times 10^{-3}$ & 
     $(3.53\,\pm 0.21)\times 10^{-3}$ \\
$D^0\ra\eta\pi^0$  & $(0.67\,\pm 0.10\,\pm 0.05)\times 10^{-3}$ & 
     $(0.68\,\pm 0.07)\times 10^{-3}$ \\
\hline
\end{tabular}
\caption{BESIII measurements of singly Cabibbo-suppressed decays.}
\label{tab:hadronic_besiii}
\end{center}
\end{table}

\subsection{BaBar}

\begin{figure}[htb]
\vskip-0.8in
\centering
\includegraphics[width=100mm,angle=-90]{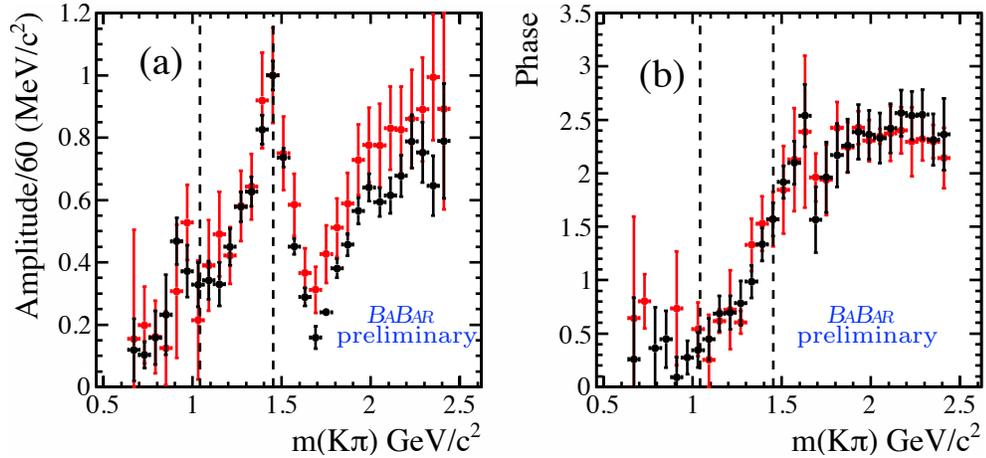}
\vskip-0.8in
\caption{BaBar results for (a) magnitudes and (b) phases
in bins of $M^{}_{K\pi}$ from a Dalitz analysis of 
$\eta^{}_c\ra K^+ K^0_S\,\pi^-$ (black) and
$\eta^{}_c\ra K^+ K^-\pi^0$ (red) decays.
The dashed lines indicate the $K \eta$ and $K \eta'$ thresholds.
The distributions indicate the presence of
the $K^*(1430)$ resonance. }
\label{fig:dalitz_babar}
\end{figure}

BaBar (Palano) reported results of a model-independent Dalitz analysis
of four decay modes: 
$\eta^{}_c\ra K^+ K^0_S\,\pi^-$,
$\eta^{}_c\ra K^+ K^-\pi^0$,
$J/\psi\ra \pi^+\pi^-\pi^0$, and
$J/\psi\ra K^+ K^-\pi^0$.
For all modes an unbinned likelihood fit is performed
for magnitudes and phases in 30 bins of $K\pi$ or $\pi\pi$
invariant mass. Both $\eta^{}_c\ra K^+ K \pi$ samples show
clear evidence for the $K^{*0}(1430)$ resonance: 
at this mass value the magnitude reaches a maximum and the
phase passes through 90$^\circ$ (see Fig.~\ref{fig:dalitz_babar}).
This behavior differs somewhat from that observed in
a model-independent Dalitz analysis of $D^+\ra K^+\pi^+\pi^-$ 
decays by FNAL E791~\cite{e791_kpi_modelind}.

BaBar also reported a search for anomalous structure (Sokoloff).
A sample of $B$ decays was fully reconstructed, and, among the
remaining tracks, a $K^+$ was required and its momentum plotted.
A sharp peak in this spectrum would indicate a two-body
decay $B\ra K^+X$ of the other $B$ in the event.
Both $B^0$ and $B^+$ samples were analyzed, and several
momentum peaks were observed corresponding to decays
$B\ra DK^+$, $B\ra D^*K^+$, and $D^{**}(2420)K^+$.
A sharp peak was observed in the $B^+$ sample corresponding
to $B^+\ra \overline{D}^{\,**0}(2680)K^+$, but no analogous
peak was seen in the $B^0$ sample.

\subsection{LHCb}

LHCb (Palano again) presented Dalitz analyses for the modes
$B^-\ra D^+ K^-\pi^-$, 
$B^0\ra \dbar \pi^+\pi^-$, 
$B^0\ra \dbar K^+\pi^-$, and
$B^0_s\ra \dbar K^-\pi^+$.
These analyses fit the data using isobar models.
For $B^-\ra D^+ K^-\pi^-$, the well-known resonances 
$D^*_0(2400)^0$, $D^*_2(2460)^0$, and $D^*_1(2760)^0$
are observed. The resulting fit fractions are listed in
Table~\ref{tab:dalitz_lhcb}.

\begin{table}[hbt]
\renewcommand{\arraystretch}{1.1}
\begin{center}
\begin{tabular}{lc} 
\hline 
 Resonance & Fit fraction \\ 
 \hline \\ [-2.5ex] 
$D^{*}_{0}(2400)^{0}$ & $\phantom{2}8.3 \pm 2.6 \pm 0.6 \pm \phantom{2}1.9$ \\ 
$D^{*}_{2}(2460)^{0}$ & $31.8 \pm 1.5 \pm 0.9 \pm \phantom{2}1.4$ \\ 
$D^{*}_{1}(2760)^{0}$ & $\phantom{2}4.9 \pm 1.2 \pm 0.3 \pm \phantom{2}0.9$ \\ 
\hline 
$S$-wave non-resonant & $38.0 \pm 7.4 \pm 1.5 \pm 10.8$ \\ 
$P$-wave non-resonant & $23.8 \pm 5.6 \pm 2.1 \pm \phantom{2}3.7$ \\ 
\hline \\ [-2.5ex] 
$D^{*}_{v}(2007)^{0}$ & $\phantom{2}7.6 \pm 2.3 \pm 1.3 \pm \phantom{2}1.5$ \\ 
$B^*_v$ & $\phantom{2}3.6 \pm 1.9 \pm 0.9 \pm \phantom{2}1.6$ \\ 
\hline 
\end{tabular} 
\caption{\small LHCb results for fitted fractions (\%)
in $B^-\ra D^+K^-\pi^-$ decays. The errors listed are
statistical, experimental systematic, and model
uncertainties, respectively~\cite{dalitz_lhcb}.}
\label{tab:dalitz_lhcb}
\end{center}
\end{table}

LHCb also reported a measurement of 
$B^+\ra X(3872) K^+,\,X(3872)\ra J/\psi\,\rho^0$ decays.
The experiment reconstructs 1011 candidates with a purity
of 80\%. By fitting the helicity angle distribution, LHCb
confirms that the $X(3872)$ has quantum numbers $J^{PC}=1^{++}$.

\section{\boldmath Semileptonic Decays}

\subsection{BESIII}

BESIII (An, Ma) presented results for several semileptonic decays.
For $D^+\ra K^-\pi^+ e^+\nu$ they measure the branching fraction
to be $(3.71\,\pm 0.03\,\pm0.09)$\%. Most of the $K^-\pi^+$
pairs originate from $K^{*0}$ decays; requiring 
$M^{}_{K\pi}\in (0.8,1.0)$\gevm\ yields a branching
fraction of $(3.33\,\pm 0.03\,\pm0.08)$\%.
For $D^+\ra \omega e^+\nu$ BESIII observes a significant
signal and measures a branching fraction of 
$(1.63\,\pm 0.11\,\pm0.08)\times 10^{-3}$. The
statistics are sufficient to extract form factor parameters
for the first time; the results are
$r^{}_V\equiv V(0)/A^{}_1(0)=1.24\,\pm 0.09\,\pm0.06$ and 
$r^{}_2\equiv A^{}_2(0)/A^{}_1(0)=1.06\,\pm 0.15\,\pm0.05$.
Finally, results for $D^+\ra K^{}_L e^+\nu$ were presented.
The branching fraction is measured to be
$(4.482\,\pm 0.027\,\pm0.103)$\%, and the form factor
parameters are
$|V^{}_{cs}| f^K_+(0) = 0.728\,\pm 0.006\,\pm0.011$ and 
$r^{}_1\equiv a^{}_1/a^{}_0 = 1.91\,\pm 0.33\,\pm0.24$.
As this final state is self-tagging, signal events are
subsequently divided into $e^+$ and $e^-$ subsamples
and the \cp\ asymmetry measured. The result is
$A^{K^{}_L e\nu}_{CP} = (-0.59\,\pm 0.60\,\pm 1.50)$\%.

\subsection{BaBar}

BaBar (Oyanguren) measures the branching fraction for $D^0\ra \pi^- e^+\nu$
decays normalized to $D^0\ra K^-\pi^+$ decays. To reduce backgrounds
they require that the $D^0$ originate from $D^{*+}\ra D^0\pi^+_s$
decays. The signal yield is obtained by fitting the
$\Delta M = M^{}_{(\pi e\nu)\pi^+_s}-M^{}_{\pi e\nu}$ distribution.
The resulting ratio of branching fractions is
\begin{eqnarray}
R^{}_D & \equiv & 
\frac{{\cal B}(D^0\ra\pi^- e^+\nu)}{{\cal B}(D^0\ra K^-\pi^+)}
\ =\ 0.0702\,\pm 0.0017\,\pm 0.0023\,.
\end{eqnarray}
Using the world average value for ${\cal B}(D^0\ra K^-\pi^+)$
yields ${\cal B}(D^0\ra\pi^- e^+\nu)= 
(2.770\,\pm 0.068\,\pm 0.092\,\pm 0.037)\times 10^{-3}$,
where the last error is from ${\cal B}(D^0\ra K^-\pi^+)$. 
BaBar subsequently fits the $z$(-expansion) distribution
to measure the normalization factor
$|V^{}_{cd}| f^{\pi}_+(0) = 
0.1374\,\pm0.0038\,\pm0.0022\,\pm0.0009^{}_{\rm ext}$,
where the last error is due to external factors not directly
related to the experimental measurement.
Inserting $|V^{}_{cd}|=|V^{}_{us}|=0.2252\,\pm0.0009$~\cite{PDG13}
gives a form factor normalization
$f^{\pi}_+(0) = 0.610\,\pm0.017\,\pm0.010\,\pm0.005^{}_{\rm ext}$.
Alternatively, inserting the average form factor from lattice QCD (LQCD)
calculations, $f^{\pi}_+(0) = 0.666\,\pm0.029$~\cite{lattice_avg},
gives $|V^{}_{cd}| = 0.206\,\pm0.007^{}_{\rm exp}\,\pm0.009^{}_{\rm LQCD}$.
This value is consistent within errors with the current world
average as calculated by HFAG using  $D\ra\pi\ell\nu$ and
$D\ra\ell\nu$ decays: $0.219\pm 0.006$~\cite{hfag_Vcd}.

\section{\boldmath Leptonic Decays}

BESIII (Ma) presented results for purely leptonic
$D^+\ra e^+\nu,\,\mu^+\nu,\,\tau^+\nu$ decays, and
Belle (Eidelman) presented results for leptonic 
$D^+_s\ra e^+\nu,\,\mu^+\nu,\,\tau^+\nu$ decays.
The branching fractions are used to determine the products
of decay constants multiplied by CKM matrix elements
$f^{}_D|V^{}_{cd}|$ and $f^{}_{D^{}_s}|V^{}_{cs}|$, respectively.
The Belle results are the world's most precise. 

It is notable that the current world average for 
$|V^{}_{cd}|$~\cite{hfag_Vcd} is dominated by measurements
of purely leptonic $D^+\ra e^+\nu,\,\mu^+\nu,\,\tau^+\nu$ decays:
$0.219\,\pm0.005\,\pm0.003$. This value has a smaller overall
error than that obtained from semileptonic $D^0\ra\pi^-\ell^+\nu$
decays, $0.214\,\pm0.003\,\pm0.009$. Similarly, the current
world average for $|V^{}_{cs}|$~\cite{hfag_Vcs} ($0.998\,\pm0.020$)
is dominated by measurements of leptonic
$D^+_s\ra e^+\nu,\,\mu^+\nu,\,\tau^+\nu$ decays,
$1.008\,\pm0.018\,\pm0.011$, i.e., the overall error is
(slightly) smaller than that obtained from semileptonic
$D^0\ra K^-\ell^+\nu$ decays, $0.975\,\pm0.007\,\pm0.025$.

\section{\boldmath Rare, Forbidden, and Radiative Decays}

\subsection{BESIII}

BESIII (Zhao) presented results for flavor-changing neutral-current
(FCNC) and lepton-number-violating (LNV) decays of $D^+_{(s)}$ mesons
involving electrons; these are summarized in Table~\ref{tab:fcnc_besiii}.
In all cases no signal was observed and upper limits were set.
For $D^+\ra K^- e^+ e^+$ and $D^+_s\ra\pi^+ e^+ e^-$, these
limits are the world's most stringent.
BESIII also obtained an upper limit for the purely radiative
decay $D^0\ra\gamma\gamma$ (Table~\ref{tab:fcnc_besiii});
however, this limit is almost twice the corresponding
upper limit set by BaBar~\cite{gammagamma_babar}.

\begin{table}[hbt]
\renewcommand{\arraystretch}{1.1}
\begin{center}
\begin{tabular}{l|cc}  
Mode &  90\% C.L. upper limit & PDG upper limit \\
\hline
$D^+\ra K^+ e^+ e^-$    & $1.2\times 10^{-6}$ & $1.0\times 10^{-6}$ \\
$D^+\ra K^- e^+ e^+$    & $0.6\times 10^{-6}$ & $0.9\times 10^{-6}$ \\
$D^+_s\ra\pi^+ e^+ e^-$ & $0.3\times 10^{-6}$ & $1.1\times 10^{-6}$ \\
$D^+_s\ra\pi^- e^+ e^+$ & $1.2\times 10^{-6}$ & $1.1\times 10^{-6}$ \\
\hline
$D^0\ra\gamma\gamma$ & $3.8\times 10^{-6}$ & $2.2\times 10^{-6}$ \\
\hline
\end{tabular}
\caption{90\% C.L. upper limits from BESIII for FCNC and LNV decays,
and for the radiative decay $D^0\ra\gamma\gamma$~\cite{gammagamma_bes3}.}
\label{tab:fcnc_besiii}
\end{center}
\end{table}

\subsection{LHCb}

LHCb (G\"{o}bel, Vacca) presented a half dozen results for
FCNC and LNV decays involving muons; these are summarized in
Table~\ref{tab:lhcb_fcnc}. To reduce backgrounds the $D^0$ is
required to originate from $D^{*+}\ra D^0\pi^+$ decays.
In all cases no signal was observed and upper limits were set.
The limits for $D^0\ra\mu^+\mu^-$ and $D^0\ra\pi^+\pi^-\mu^+\mu^-$
are about two orders of magnitude larger than the expected Standard
Model (SM) rate. These analyses all have substantial backgrounds from
hadronic $D\ra\pi^+\pi^- n(\pi)$ decays in which $\pi^+\ra\mu^+\nu$,
and these backgrounds produce mass peaks that overlap with those of
the signal. LHCb is able to discriminate these backgrounds from
signal due to their high statistics.

\begin{table}[hbt]
\renewcommand{\arraystretch}{1.1}
\begin{center}
\begin{tabular}{l|c}  
Mode &  90\% C.L. upper limit \\
\hline
$D^0\ra\mu^+\mu^-$ & $6.2\times 10^{-9}$ \\
$D^0\ra\pi^+\pi^-\mu^+\mu^-$ & $5.5\times 10^{-7}$ \\
\hline
$D^+\ra\pi^+\mu^+\mu^-$ & $7.3\times 10^{-8}$ \\
$D^+\ra\pi^-\mu^+\mu^+$ & $2.2\times 10^{-8}$ \\
\hline
$D^+_s\ra\pi^+\mu^+\mu^-$ & $4.1\times 10^{-7}$ \\
$D^+_s\ra\pi^-\mu^+\mu^+$ & $1.2\times 10^{-7}$ \\
\hline
\end{tabular}
\caption{90\% C.L. upper limits from LHCb for FCNC and LNV decays.}
\label{tab:lhcb_fcnc}
\end{center}
\end{table}

\section{\boldmath $T$ Violation}

\begin{figure}[htb]
\vskip-0.4in
\centering
\includegraphics[width=44mm,angle=-90]{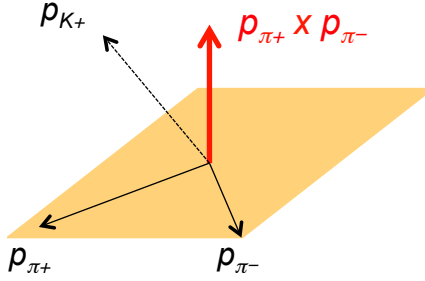}
\caption{$T$-violating observable
$p^{}_{K^+}\cdot (p^{}_{\pi^+}\times p^{}_{\pi^-})$}
\label{fig:tviol}
\end{figure}

LHCb (Martinelli) presented a measurement of $T$ violation in
$D^0\ra K^+K^-\pi^+\pi^-$ decays. The $D^0$ is required to
originate from $D^{*+}\ra D^0\pi^+_s$, and the charge of the
$\pi^\pm_s$ is used to divide signal events into $D^0$ and $\dbar$
subsamples. For these subsamples LHCb calculates the observables
\begin{eqnarray}
C^{}_T & = & p^{}_{K^+}\cdot (p^{}_{\pi^+}\times p^{}_{\pi^-})
\hskip 0.30in (D^0 {\rm\ decays}) \\
\overline{C}^{}_T & = & p^{}_{K^-}\cdot (p^{}_{\pi^-}\times p^{}_{\pi^+})
\hskip 0.30in (\dbar {\rm\ decays})\,.
\end{eqnarray}
These observables represent the projection of the $K^+$ or $K^-$
momentum onto the normal to the $(\pi^+, \pi^-)$ decay plane; 
see Fig.~\ref{fig:tviol}. Under
a $T$ transformation particle momenta are reversed, and
$C^{}_T = -\overline{C}^{}_T$. To quantify a deviation
from this equality, one constructs two variables
\begin{eqnarray}
A^{}_T & = & \frac{\Gamma(C^{}_T>0) - \Gamma(C^{}_T<0)}{\Gamma} \\
\overline{A}^{}_T & = & 
\frac{\Gamma(-\overline{C}^{}_T>0) - \Gamma(-\overline{C}^{}_T<0)}{\overline{\Gamma}}\,;
\end{eqnarray}
the measure of $T$ violation is then
$a^{}_T = (A^{}_T - \overline{A}^{}_T)/2$. In general, 
$A^{}_T, \overline{A}^{}_T\neq 0$
due to resonant structure in the 4-body final state.
All available measurements of $a^{}_T$ are tabulated by
the Heavy Flavor Averaging Group (HFAG) in Ref.~\cite{hfag_tviol}.

The LHCb results are:
\begin{eqnarray}
A^{}_T(D^0) & = & (-71.8\,\pm 4.1\,\pm 1.3)\times 10^{-3} \\
\overline{A}^{}_T(\dbar) & = & (-75.5\,\pm 4.1\,\pm 1.2)\times 10^{-3} \\
\Rightarrow \hskip0.20in a^{}_T & = & 
(1.8\,\pm 2.9\,\pm 0.4)\times 10^{-3}\,.
\end{eqnarray}
Thus they see no evidence of $T$-violation. LHCb also 
calculates $a^{}_T$ in 32 separate bins of phase space and in
four bins of $D^0$ decay time. In both cases all values are
consistent with zero. Fitting these values to a constant
$a^{}_T\!=\!0$ yields a $p$-value of 0.74 for the set of phase space
measurements, and $p=0.83$ for the set of decay time measurements.

\section{\boldmath Mixing and $CP$ Violation}

At this workshop, three recent measurements of mixing and
\cp\ violation (\cpv) parameters in the $D^0$-$\dbar$ system
were presented:
\ycp\ by BESIII, \agamma\ by CDF, and \agamma\ by LHCb.
These measurements are used by HFAG in their global fit for
mixing and \cpv\ parameters $x,\,y,\,|q/p|$ and $\phi$, and
are discussed below.
In addition, CLEOc (Libby) showed evidence that the decay
$D^0\ra\pi^+\pi^-\pi^0$ is \cp-even. 

\subsection{BESIII}

BESIII (Albayrak) measures 
$y^{}_{CP}\equiv (\Gamma^{}_{CP+} - \Gamma^{}_{CP-})/2\Gamma$,
where $\Gamma = (\Gamma^{}_{CP+} + \Gamma^{}_{CP-})/2$,
using semileptonic decays. Inverting this relation gives 
$\Gamma^{}_{CP\pm} = \Gamma(1\pm y^{}_{CP})$, and thus
\begin{eqnarray}
{\cal B}(CP+\,\ra K\ell\nu) & = &   
\frac{\Gamma^{}_{CP+\,\ra K\ell\nu}}{\Gamma^{}_{CP+}}
\ =\ \frac{\Gamma^{}_{CP+\,\ra K\ell\nu}}{\Gamma (1+y^{}_{CP})}
\ \approx\ \frac{\Gamma^{}_{CP+\,\ra K\ell\nu}}{\Gamma}(1-y^{}_{CP})\,.
\end{eqnarray}
Similarly, 
${\cal B}(CP-\ra K\ell\nu)\approx 
(\Gamma^{}_{CP-\,\ra K\ell\nu}/\Gamma)\times (1+y^{}_{CP})$.
Neglecting \cpv\ in $D^0$ decays, 
$\Gamma^{}_{CP+\,\ra K\ell\nu} = \Gamma^{}_{CP-\,\ra K\ell\nu}$
and one can show
\begin{eqnarray}
y^{}_{CP} & \approx  & \frac{1}{4}
\left( \frac{{\cal B}(CP-\ra K\ell\nu)}{{\cal B}(CP+\ra K\ell\nu)} -
\frac{{\cal B}(CP+\ra K\ell\nu)}{{\cal B}(CP-\ra K\ell\nu)}\right)\,.
\end{eqnarray}
This method to determine $y^{}_{CP}$ is advantageous as many
systematic errors cancel in the ratio of branching fractions.
The \cp\ of the decaying $D^0$ or $\dbar$ is identified by
reconstructing the opposite-side $D$ decay in a \cp-specific
final state. The \cp-even final states used are
$K^+K^-,\,\pi^+\pi^-$, and $K^0_S\,\pi^0\pi^0$; 
the \cp-odd final states used are
$K^0_S\,\pi^0,\,K^0_S\,\omega$, and $K^0_S\,\eta$; and
the semileptonic final states used are $K^\pm\mu^\mp\nu$
and $K^\pm e^\mp\nu$. The result of the measurement
is $y^{}_{CP} = (-2.0\,\pm 1.3\,\pm 0.7)$\%.
Although this result is less precise than that of 
other experiments using hadronic decays, it is the
first such measurement using semileptonic decays,
and the precision should improve with more data.

\subsection{CDF}

CDF (Leo) analyzes \cp-even hadronic decays
$D^0\ra K^+ K^-,\,\pi^+\pi^-$ to measure the \cp-violating
parameter~\agamma.
The observable used is the time-dependent \cp\ asymmetry
\begin{eqnarray}
A^{}_{CP}(t) & \equiv & \frac{N^{}_{D^0}(t) - N^{}_{\dbar}(t)}
{N^{}_{D^0}(t) + N^{}_{\dbar}(t)}\ \approx\ 
A(0) - A^{}_\Gamma\left(\frac{t}{\tau}\right)\,,
\end{eqnarray}
where the intercept $A(0)$ may be nonzero due to 
possible direct \cpv\ in the decay amplitude
and any production or reconstruction asymmetries.
By fitting the $A^{}_{CP}(t)$ distribution to determine
its slope, one obtains \agamma. The CDF results are
\begin{eqnarray}
A^{}_\Gamma(KK) & = & (-1.9\,\pm 1.5\,\pm 0.4)\times 10^{-3} \\
A^{}_\Gamma(\pi\pi) & = & (-0.1\,\pm 1.8\,\pm 0.3)\times 10^{-3} \\
\langle A^{}_\Gamma \rangle & = & (-1.2\,\pm 1.2)\times 10^{-3}\,,
\end{eqnarray}
where the last result is a weighted average of the $K^+K^-$
and $\pi^+\pi^-$ results.

\subsection{LHCb}

\begin{figure}[htb]
\vskip-1.2in
\centering
\includegraphics[width=115mm,angle=-90]{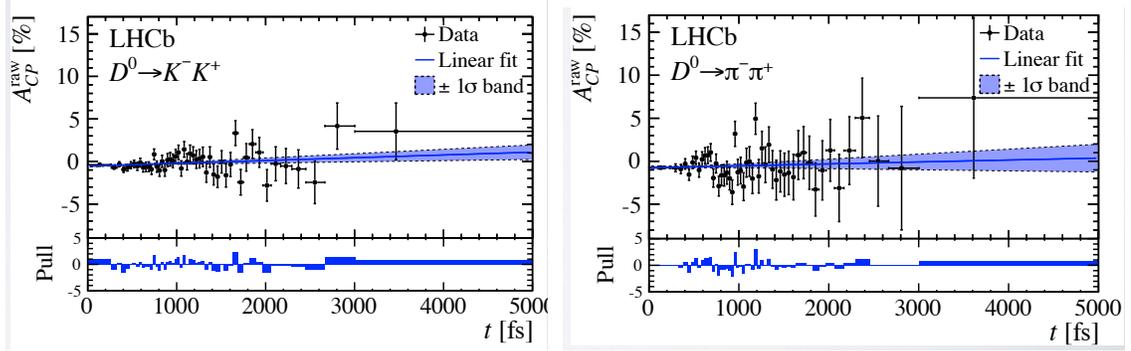}
\vskip-1.3in
\caption{LHCb fit to the $D^0$ decay time distribution to
determine $A^{}_\Gamma$, using $D^0\ra K^+K^-$ (left) and
$D^0\ra\pi^+\pi^-$ (right) decays.}
\label{fig:agamma_lhcb}
\end{figure}

LHCb (Reichert, Naik) measures \agamma\ using the same method as
that used by CDF. The fit to the data is shown in Fig.~\ref{fig:agamma_lhcb}.
Due to the high statistics of the LHCb dataset, these results
are the world's most precise:
\begin{eqnarray}
A^{}_\Gamma(KK) & = & (-1.34\,\pm 0.77\,^{+0.26}_{-0.34})\times 10^{-3} \\
A^{}_\Gamma(\pi\pi) & = & (-0.92\,\pm 1.45\,^{+0.25}_{-0.33})\times 10^{-3}\,.
\end{eqnarray}
Taking a weighted average of these values assuming
the errors are uncorrelated gives
\begin{eqnarray}
\langle A^{}_\Gamma \rangle & = & (-1.24\,^{+0.71}_{-0.73})\times 10^{-3}\,.
\end{eqnarray}

\subsection{Heavy Flavor Averaging Group Results}

\subsubsection{Global fit}

HFAG performs a global fit to 45 observables measured 
from \dklnu, \dkk, \dpipi, \dkpi, $D^0\ra K^+\pi^-\pi^0$, 
\dkspp, and \dkskk\ decays, and from double-tagged branching 
fractions measured at the $\psi(3770)$ resonance.
There are ten fitted parameters:
mixing parameters $x$ and $y$; 
indirect \cpv\ parameters $|q/p|$ and $\phi$; 
the ratio of decay rates
$R^{}_D\equiv
[\Gamma(D^0\ra K^+\pi^-)+\Gamma(\dbar\ra K^-\pi^+)]/
[\Gamma(D^0\ra K^-\pi^+)+\Gamma(\dbar\ra K^+\pi^-)]$;
direct \cpv\ parameters $A^{}_K$, $A^{}_\pi$, and
$A^{}_D =(R^+_D-R^-_D)/(R^+_D+R^-_D)$, where the $+\,(-)$
superscript corresponds to $D^0\,(\dbar)$ decays;
the strong phase difference
$\delta$ between $\dbar\ra K^-\pi^+$ and 
$D^0\ra K^-\pi^+$ amplitudes; and 
the strong phase difference $\delta^{}_{K\pi\pi}$ between 
$\dbar\ra K^-\rho^+$ and $D^0\ra K^-\rho^+$ amplitudes. 
The mixing parameters are defined as
$x\equiv(m^{}_1-m^{}_2)/\Gamma$ and 
$y\equiv (\Gamma^{}_1-\Gamma^{}_2)/(2\Gamma)$, where 
$m^{}_1,\,m^{}_2$ and $\Gamma^{}_1,\,\Gamma^{}_2$ are
the masses and decay widths for the $D^0$-$\dbar$
mass eigenstates, and $\Gamma =(\Gamma^{}_1+\Gamma^{}_2)/2$. 
The fitter determines central values and errors using a
$\chi^2$ statistic. Correlations among observables are
accounted for by using covariance matrices provided by
the experimental collaborations. Errors are assumed to
be Gaussian. The relationships between observables and
fitted parameters, and details of the fitting procedure,
are given in Ref.~\cite{hfag_preprint}.

\begin{figure}
\vskip-0.80in
\begin{center}
\hbox{
\includegraphics[width=76mm]{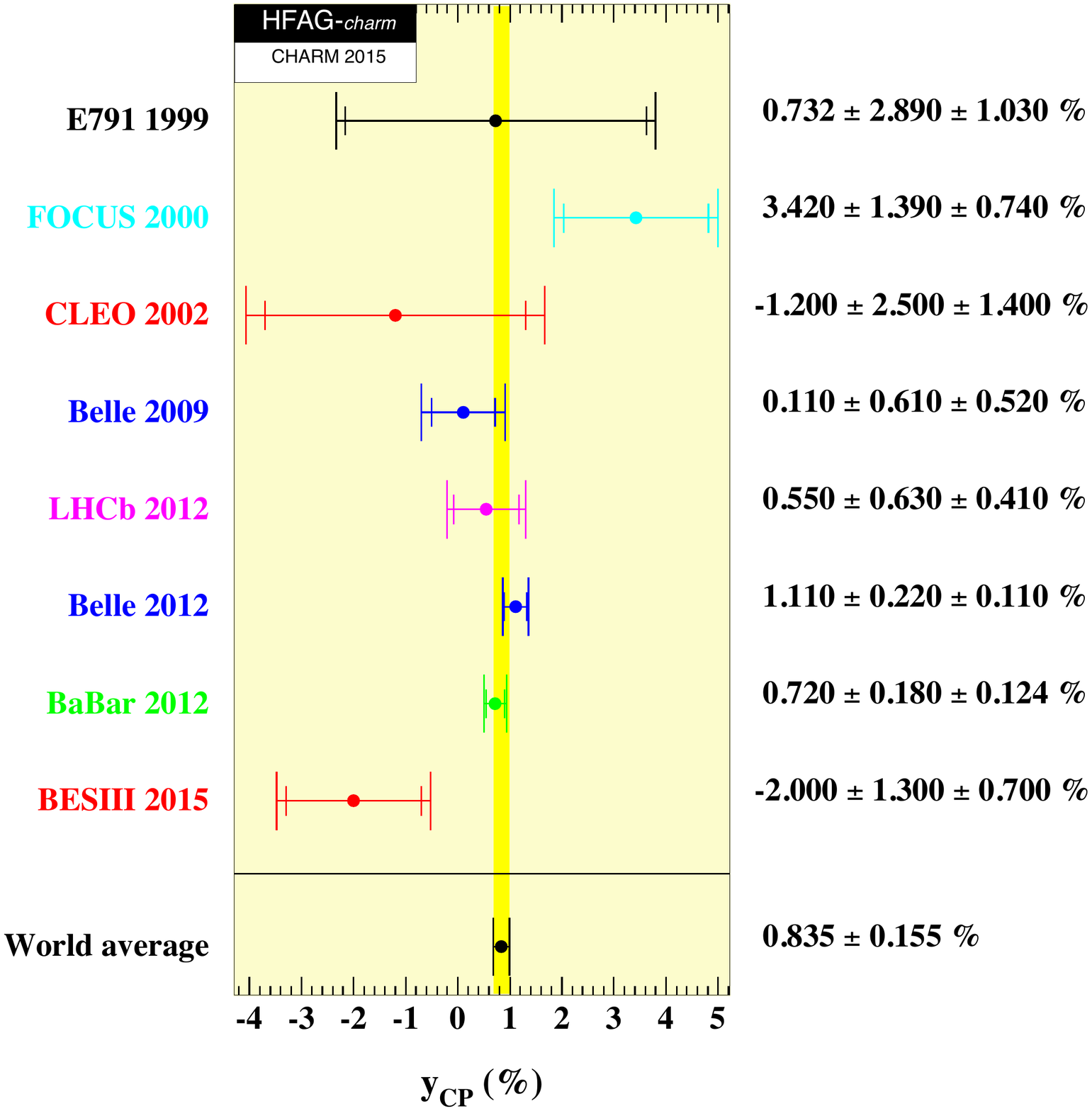}
\hskip0.10in
\includegraphics[width=76mm]{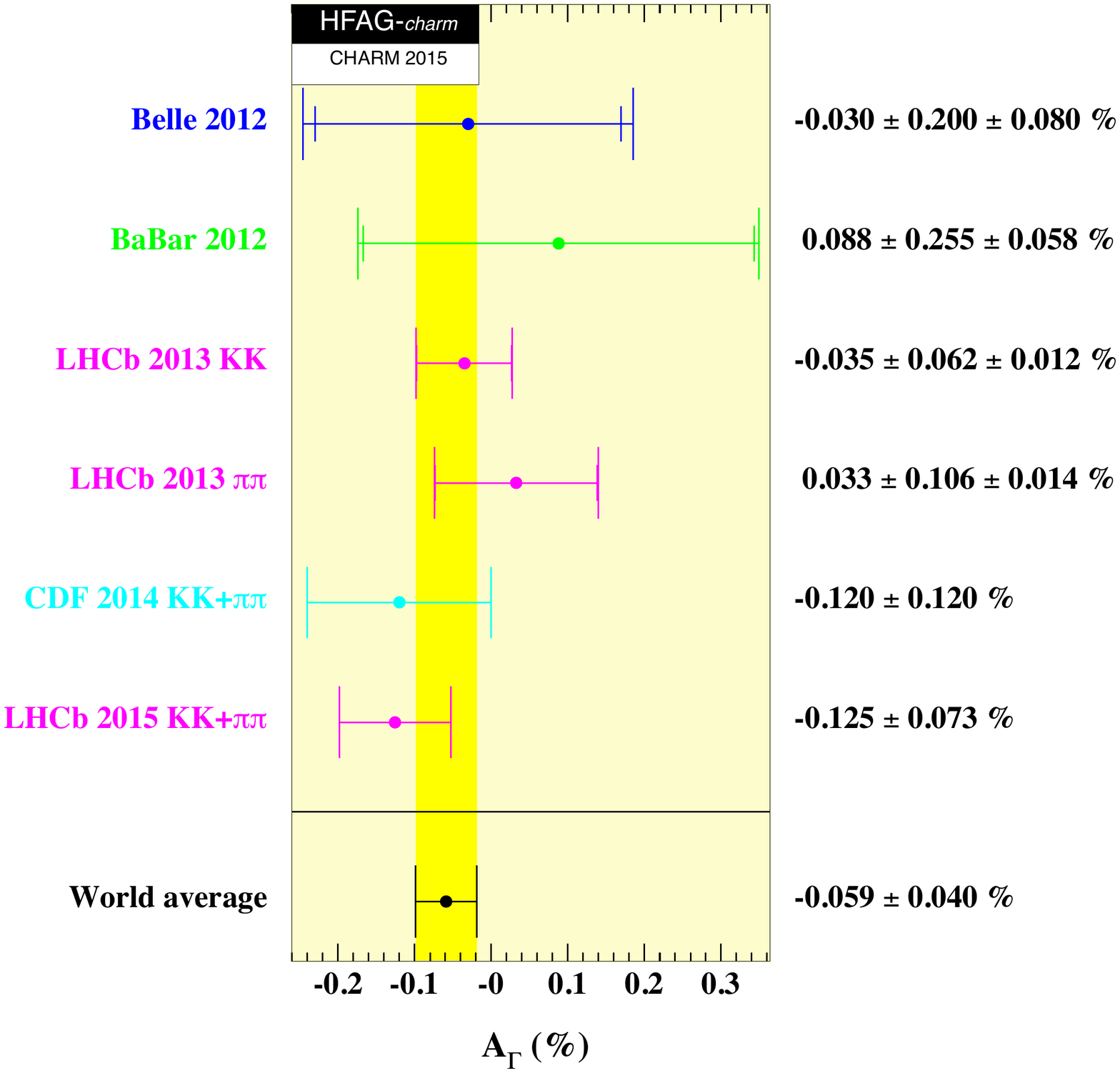}
}
\end{center}
\vskip-0.80in
\caption{\label{fig:ycp_agamma}
World average values for $y^{}_{CP}$ (left) and $A^{}_\Gamma$ (right)
as calculated by HFAG~\cite{hfag_charm15wa}. These values include
the results from BESIII and LHCb presented at this workshop.}
\end{figure}

All input measurements are given in Ref.~\cite{hfag_charm15}.
The values for observables $R^{}_M=(x^2+y^2)/2$, $y^{}_{CP}$,
and $\Gamma^{}_{CP}$ are world averages as calculated by
HFAG~\cite{hfag_charm15wa}. For the latter two observables,
the world averages used include the results presented by
BESIII and LHCb at this workshop (see Fig.~\ref{fig:ycp_agamma}).

Four types of fits are performed: 
{\it (a)}\ assuming \cp\ conservation
(fixing
$A^{}_D\!=\!0$, $A_K\!=\!0$, $A^{}_\pi\!=\!0$, $\phi\!=\!0$, 
and $|q/p|\!=\!1$);
{\it (b)}\ assuming no direct \cpv\ in doubly
Cabibbo-suppressed (DCS) decays, which fixes $A^{}_D\!=\!0$
and reduces the four independent parameters 
$(x,y,|q/p|,\phi)$ to three via the relation
$\tan\phi = (1-|q/p|^2)/(1+|q/p|^2)\times (x/y)$~\cite{Ciuchini:2007cw,Kagan:2009gb};
{\it (c)}\ same as fit {\it (b)\/} except that one fits
for alternative parameters $x^{}_{12}= 2|M^{}_{12}|/\Gamma$, 
$y^{}_{12}= \Gamma^{}_{12}/\Gamma$, and 
$\phi^{}_{12}= {\rm Arg}(M^{}_{12}/\Gamma^{}_{12})$,
where $M^{}_{12}$ and $\Gamma^{}_{12}$ are the off-diagonal
elements of the $D^0$-$\dbar$ mass and decay matrices, respectively; and
{\it (d)}\ allowing full \cpv\ (floating all parameters). 
Note that parameters $(x^{}_{12},\,y^{}_{12},\,\phi^{}_{12})$
can be derived from $(x,\,y,\,|q/p|,\,\phi)$ and
vice-versa; see Ref.~\cite{hfag_charm15}.

\begin{figure}
\vskip-0.60in
\begin{center}
\hbox{
\includegraphics[width=72mm]{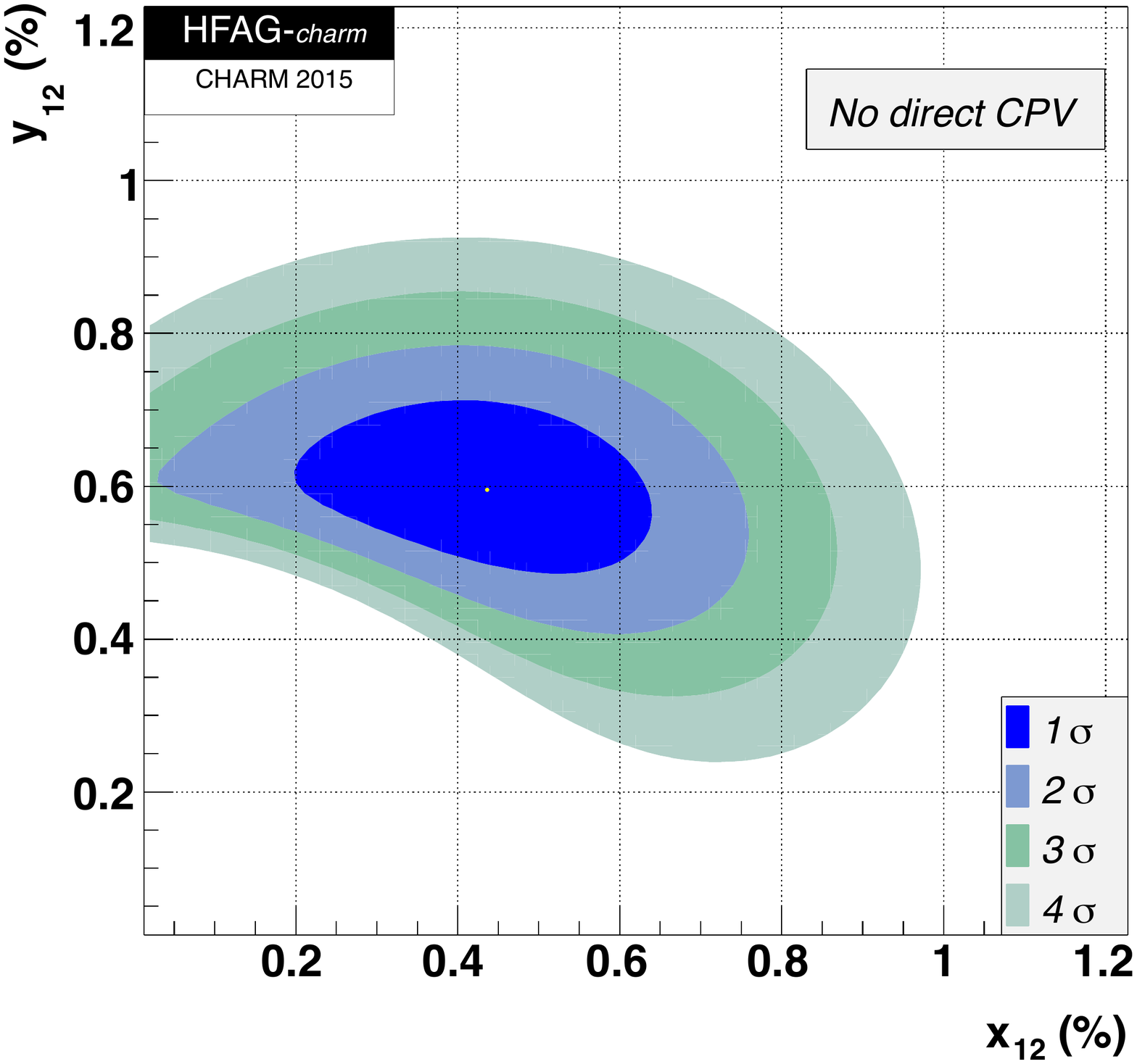}
\hskip0.10in
\includegraphics[width=72mm]{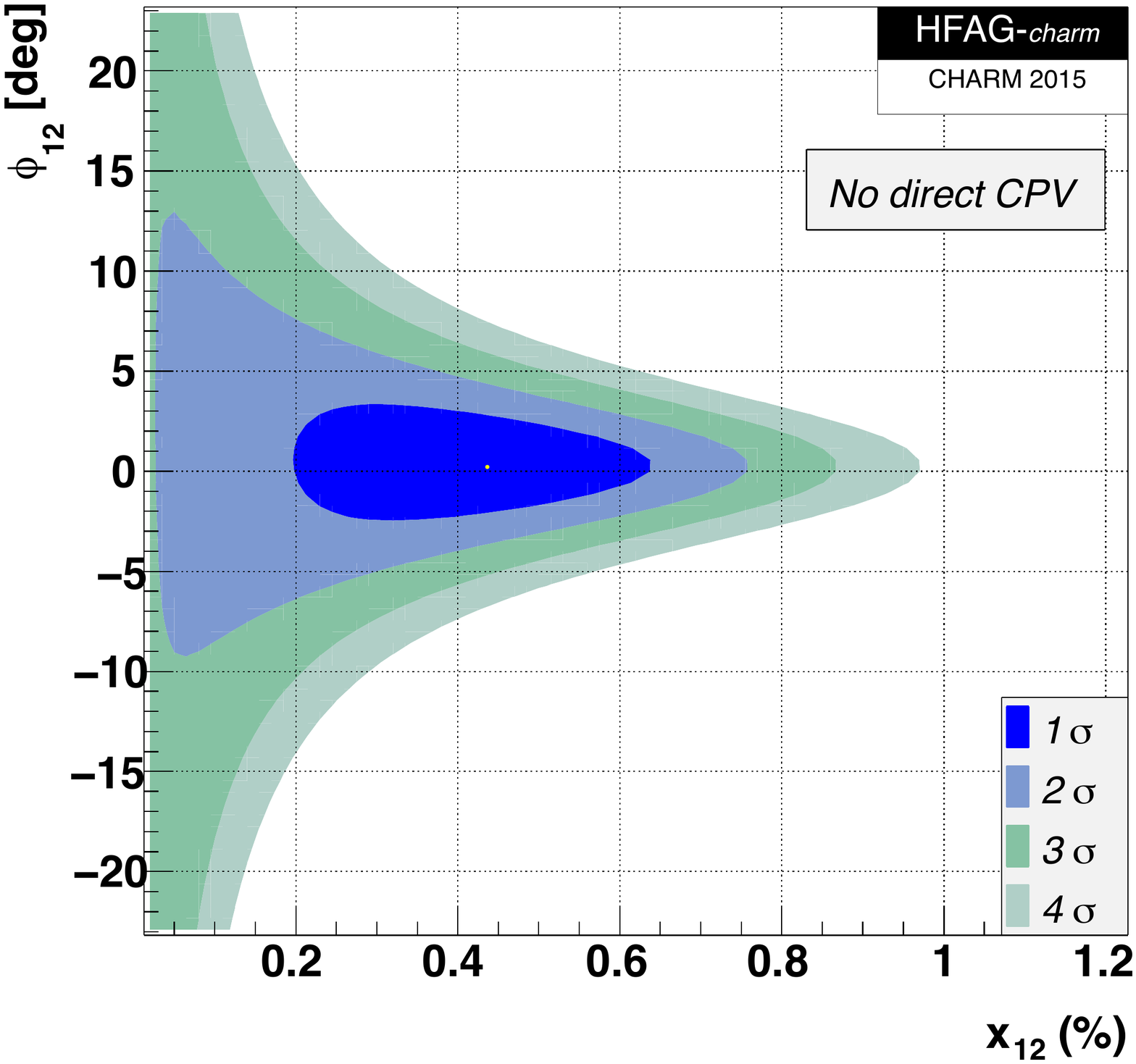}
}
\vskip-0.60in
\includegraphics[width=72mm]{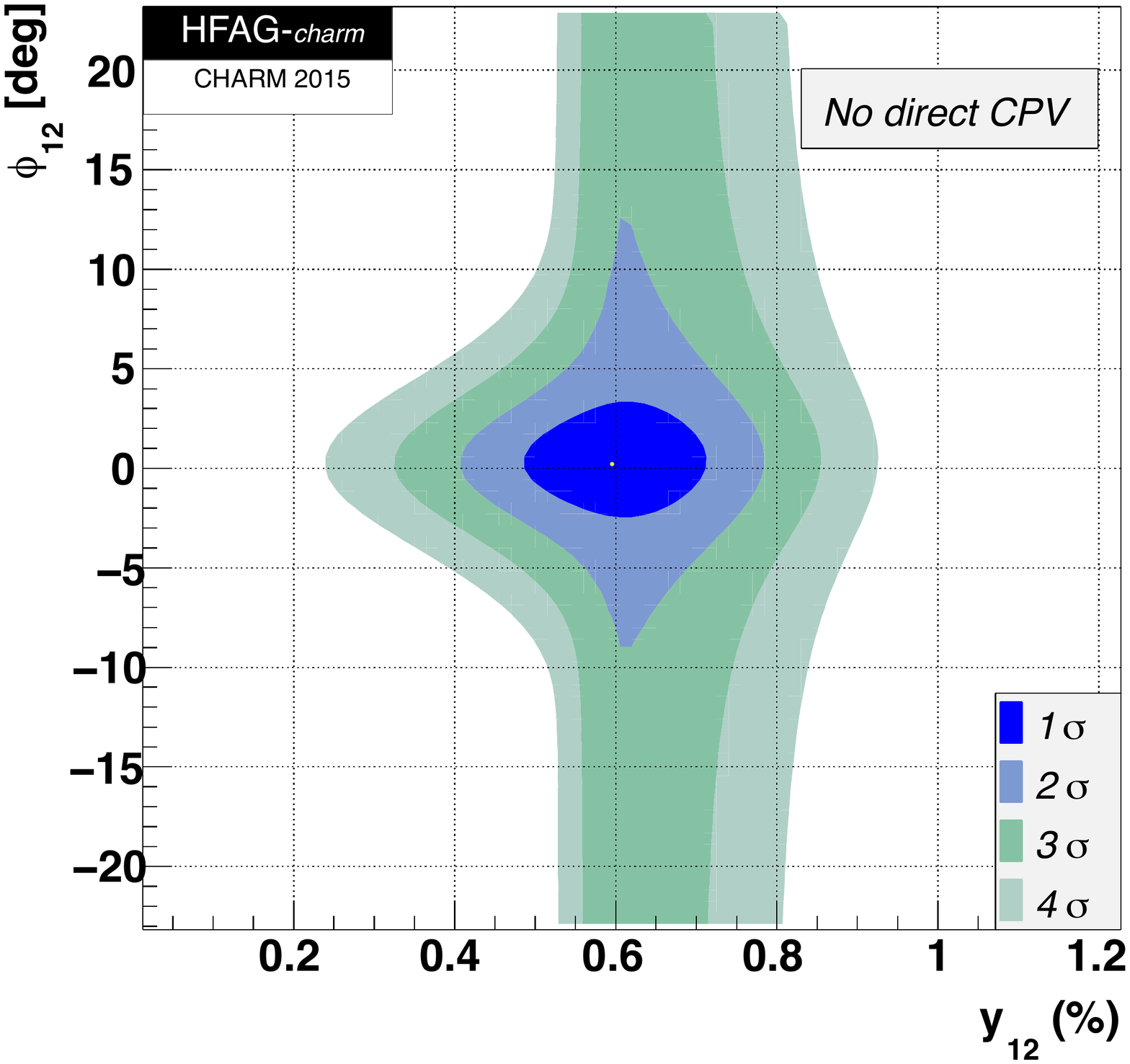}
\end{center}
\vskip-0.60in
\caption{\label{fig:contours_ndcpv}
Two-dimensional contour plots for parameters 
$(x^{}_{12},y^{}_{12})$ (top left), 
$(x^{}_{12},\phi^{}_{12})$ (top right), and 
$(y^{}_{12},\phi^{}_{12})$ (bottom) 
for no direct \cpv\ in DCS decays,
as calculated by HFAG~\cite{hfag_charm15}. }
\end{figure}

\begin{figure}
\vskip-0.80in
\begin{center}
\hbox{
\includegraphics[width=72mm]{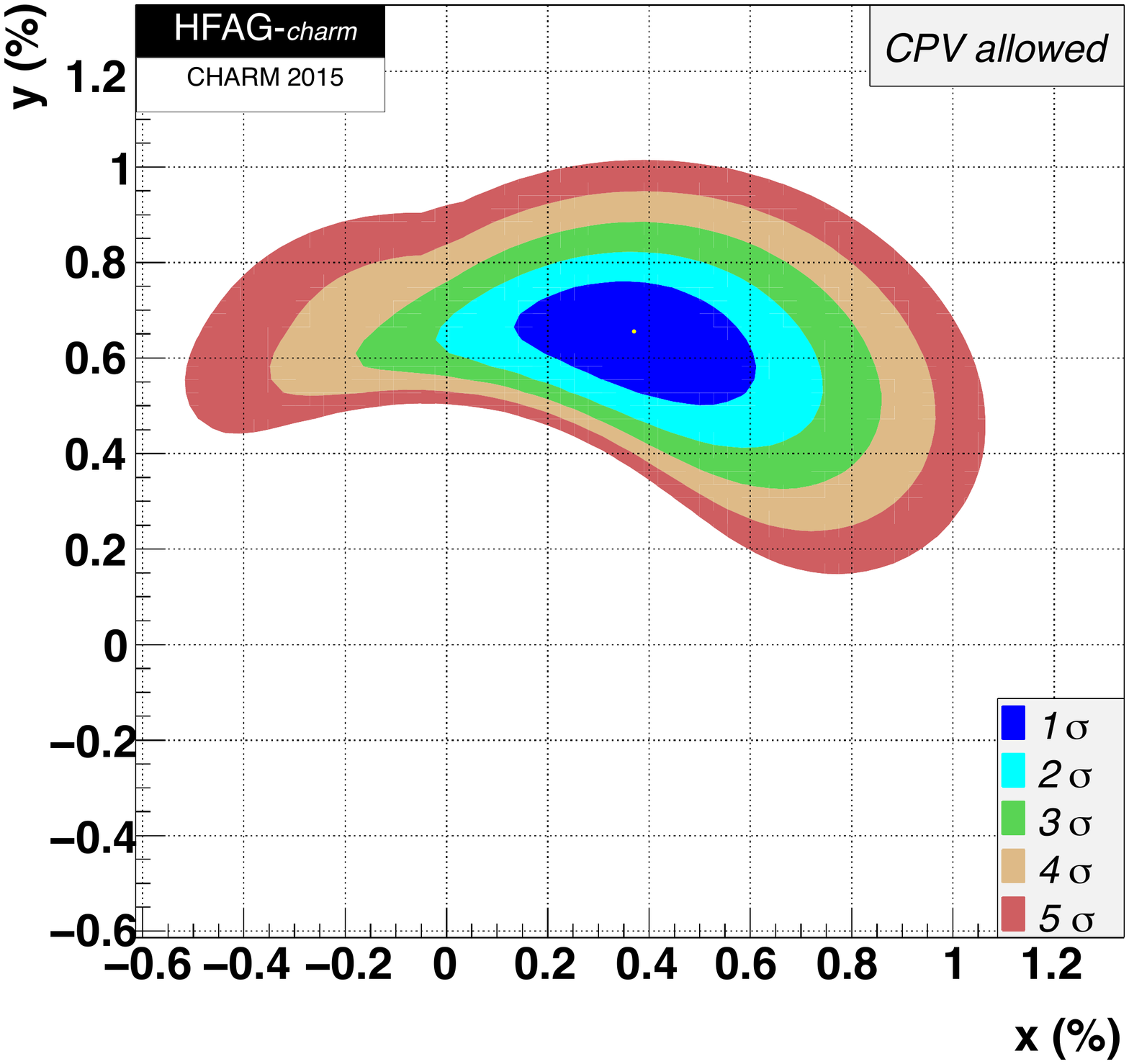}
\hskip0.10in
\includegraphics[width=72mm]{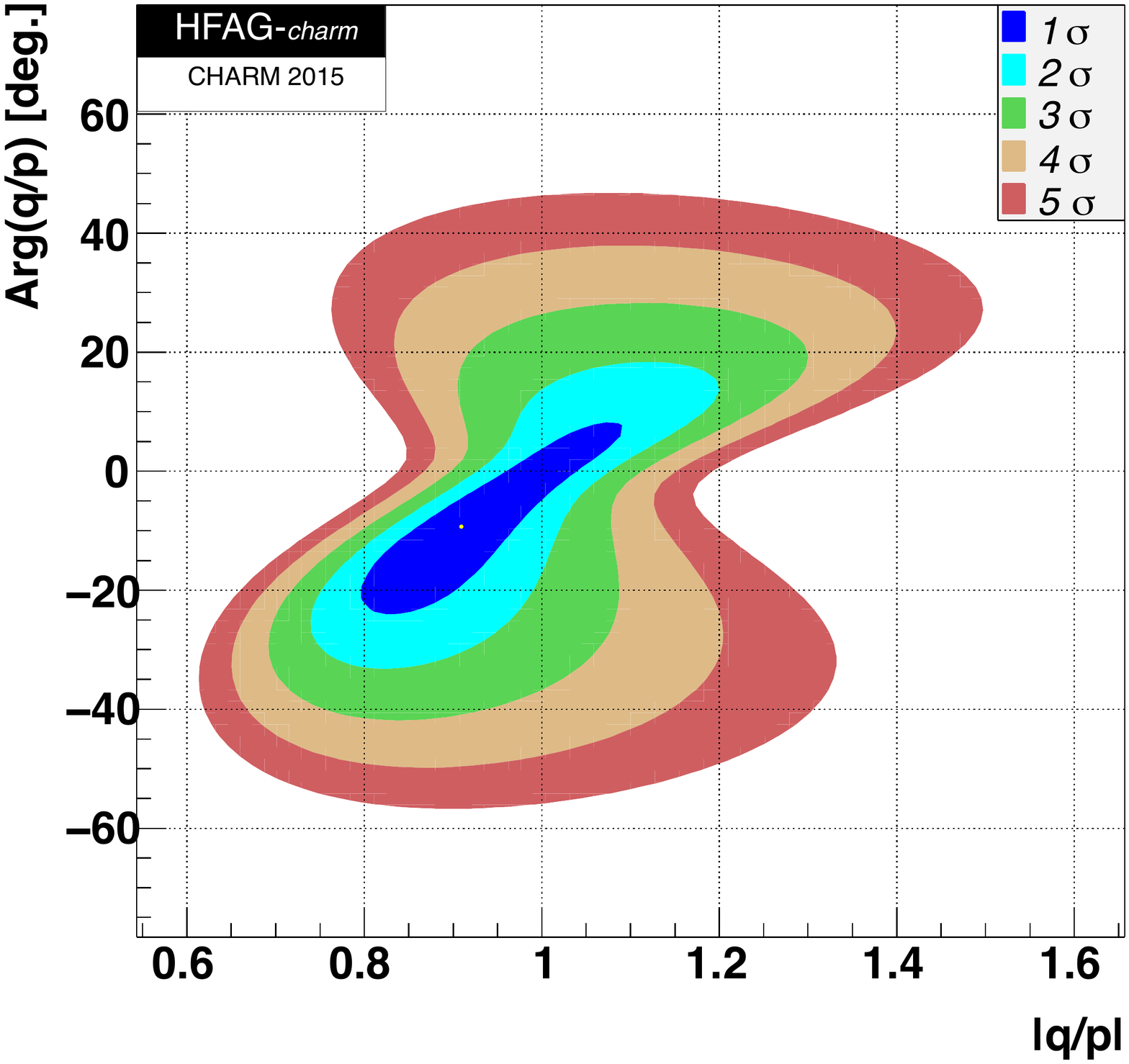}
}
\end{center}
\vskip-0.80in
\caption{\label{fig:contours_cpv}
Two-dimensional contour plots for parameters $(x,y)$ (left) 
and $(|q/p|,\phi)$ (right), allowing for \cpv,
as calculated by HFAG~\cite{hfag_charm15}. }
\end{figure}

All fit results are listed in Table~\ref{tab:hfag_global_fit},
and two-dimensional contour plots are shown
in Fig.~\ref{fig:contours_ndcpv} (no-direct-\cpv) and
Fig.~\ref{fig:contours_cpv} (all-\cpv-allowed). These
results show that $D^0$ mesons mix: 
the no-mixing point $x=y=0$ is excluded at $>12\sigma$. 
There is no evidence for \cpv\ arising from $D^0$-$\dbar$
mixing ($|q/p|\neq 1$) or from a phase difference between
the mixing amplitude and a direct decay amplitude ($\phi\neq 0$).

\begin{table}[hbt]
\renewcommand{\arraystretch}{1.2}
\begin{center}
\begin{tabular}{c|cccc}
\hline
Parameter & No \cpv & No direct \cpv & \cpv-allowed & 95\% C.L. Interval \\
 & & in DCS decays & & \\
\hline
$\begin{array}{c}
x\ (\%) \\ 
y\ (\%) \\ 
\delta^{}_{K\pi}\ (^\circ) \\ 
R^{}_D\ (\%) \\ 
A^{}_D\ (\%) \\ 
|q/p| \\ 
\phi\ (^\circ) \\
\delta^{}_{K\pi\pi}\ (^\circ)  \\
A^{}_{\pi} \\
A^{}_K \\
x^{}_{12}\ (\%) \\ 
y^{}_{12}\ (\%) \\ 
\phi^{}_{12} (^\circ)
\end{array}$ & 
$\begin{array}{c}
0.49\,^{+0.14}_{-0.15} \\
0.61\,\pm 0.08 \\
6.9\,^{+9.7}_{-11.2} \\
0.349\,\pm 0.004 \\
- \\
- \\
- \\
18.1\,^{+23.3}_{-23.8} \\
- \\
- \\
- \\
- \\
- 
\end{array}$ &
$\begin{array}{c}
0.44\,^{+0.14}_{-0.15} \\
0.60\,\pm 0.07 \\
3.6\,^{+10.4}_{-12.1} \\
0.348\,\pm 0.004 \\
- \\
1.002\,\pm 0.014 \\
-0.07\,\pm 0.6 \\
20.3\,^{+24.0}_{-24.3} \\
0.10\,\pm 0.14 \\
-0.14\,\pm 0.13 \\
0.44\,^{+0.14}_{-0.15} \\
0.60\,\pm 0.07 \\
0.2\,\pm 1.7
\end{array}$ &
$\begin{array}{c}
0.37\,\pm 0.16\\
0.66\,\,^{+0.07}_{-0.10}\\
11.8\,^{+9.5}_{-14.7} \\
0.349\,\pm 0.004 \\
-0.39\,^{+1.01}_{-1.05} \\
0.91\,^{+0.12}_{-0.08} \\ 
-9.4\,^{+11.9}_{-9.8} \\ 
27.3\,^{+24.4}_{-25.4} \\
0.10\,\pm 0.15 \\
-0.15\,\pm 0.14 \\
 \\
 \\
 \\
\end{array}$ &
$\begin{array}{c}
\left[ 0.06,\, 0.67\right] \\
\left[ 0.46,\, 0.79\right] \\
\left[ -21.1,\, 29.3\right] \\
\left[ 0.342,\, 0.357\right] \\
\left[ -2.4,\, 1.5\right] \\
\left[ 0.77,\, 1.14\right] \\\
\left[ -28.3,\, 12.9\right] \\
\left[ -23.3,\, 74.8\right] \\
\left[ -0.19,\, 0.38\right] \\
\left[ -0.42,\, 0.12\right] \\
\left[ 0.13,\, 0.69\right] \\
\left[ 0.45,\, 0.74\right] \\
\left[ -4.1,\, 4.6\right] \\
\end{array}$ \\
\hline
\end{tabular}
\caption{Results of the HFAG global fit for mixing and \cpv\ 
parameters~\cite{hfag_charm15}. The most recent results
for \ycp\ and \agamma\ are used in the fit.}
\label{tab:hfag_global_fit}
\end{center}
\end{table}

\subsubsection{Dedicated fit for {\boldmath \cpv} parameters}

HFAG also performs a fit for alternative direct and indirect \cpv\ 
parameters $a^{\rm dir}_{CP}$ and $a^{\rm ind}_{CP}$. The observables are
\begin{eqnarray}
A_{\Gamma} & \equiv & \frac{\tau(\dbar \ra h^+ h^-) - \tau(D^0 \ra h^+ h^- )}
{\tau(\dbar \ra h^+ h^-) + \tau(D^0 \ra h^+ h^- )}\,,
\end{eqnarray}
where $h^+ h^-$ is either $K^+ K^-$ or $\pi^+\pi^-$; and 
$\Delta A^{}_{CP}   \equiv A^{}_{CP}(K^+K^-) - A^{}_{CP}(\pi^+\pi^-)$,
where $A^{}_{CP}$ are time-integrated \cp\ asymmetries.
The relations between the observables and the fitted
parameters are~\cite{Gersabeck:2011xj}: 
\begin{eqnarray}
A_{\Gamma} & = & - a^{\rm ind}_{CP} - a^{\rm dir}_{CP}\,y^{}_{CP} \\ 
\Delta A^{}_{CP} & = &  \Delta a^{\rm dir}_{CP} \left( 1 + y^{}_{CP} 
\frac{\overline{\langle t\rangle}}{\tau} \right)   +   
   a^{\rm ind}_{CP} \frac{\Delta\langle t\rangle}{\tau}   +   
  \overline{a}^{\rm\,dir}_{CP}\,y^{}_{CP} \frac{\Delta\langle t\rangle}{\tau}
\nonumber \\ 
 & \approx & \Delta a^{\rm dir}_{CP}
\left( 1 + y^{}_{CP} \frac{\overline{\langle t\rangle}}{\tau} \right) 
 + a^{\rm ind}_{CP} \frac{\Delta\langle t\rangle}{\tau}\,. 
\end{eqnarray}
In the second relation, $\langle t\rangle/\tau$ denotes the mean decay 
time in units of $D^0$ lifetime; $\Delta X$ denotes the difference 
in quantity $X$ between $K^+K^-$ and $\pi^+\pi^-$ final states; and $\overline{X}$ 
denotes the average for quantity $X$. 
HFAG uses values of $\overline{\langle t\rangle}$
and $\Delta \langle t\rangle$ specific to each experiment, 
and the observables $A_\Gamma(KK)$ and $A_\Gamma(\pi\pi)$
are assumed to be identical. The measurements used and details
of the fit are given in Ref.~\cite{hfag_direct_indirect}.
Parameters $a^{\rm dir}_{CP}(K^+K^-)$ and $a^{\rm dir}_{CP}(\pi^+\pi^-)$ 
are expected to have opposite signs~\cite{Grossman:2006jg}. 

\begin{figure}
\vskip-3.7in
\centering
\hbox{
\hskip-1.0in
\includegraphics[width=195mm]{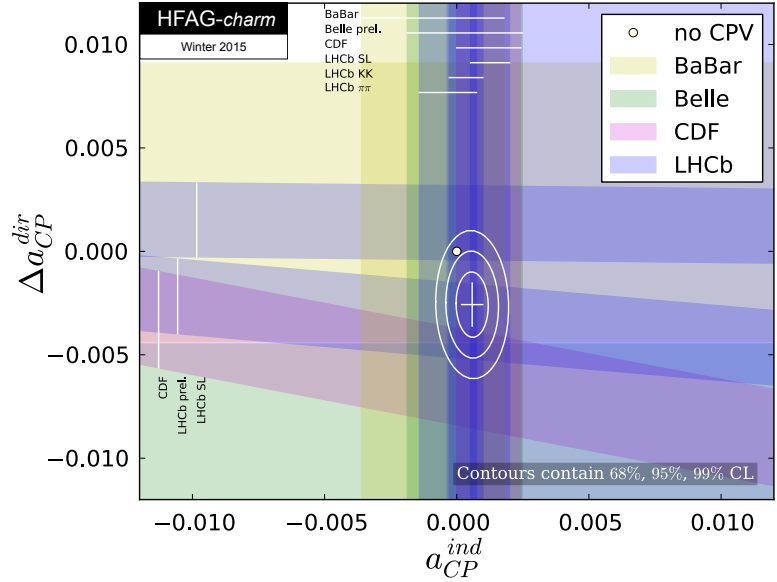}
}
\vskip-3.5in
\caption{HFAG fit for parameters $a^{\rm dir}_{CP}$ and
$a^{\rm ind}_{CP}$~\cite{hfag_direct_indirect}.
Measurements are plotted as bands showing their $\pm1\sigma$
range. The no-\cpv\ point (0,0) is shown as a filled circle,
and the best fit value is indicated by a cross showing the
one-dimensional errors.}
\label{fig:cpd_cpind_comb}
\end{figure}

The fit results are shown in Fig.~\ref{fig:cpd_cpind_comb},
which plots all relevant measurements in the two-dimensional
$(a^{\rm ind}_{CP}, \Delta a^{\rm dir}_{CP})$ plane. 
The most likely values and $\pm 1\sigma$ errors 
are~\cite{hfag_direct_indirect}:
\begin{eqnarray}
a^{\rm ind}_{CP} & = & (+0.058\pm 0.040)\% \\ 
\Delta a^{\rm dir}_{CP} & = & (-0.257\pm 0.104)\%\,.
\end{eqnarray}
Whereas $a^{\rm ind}_{CP}$ is consistent with zero,
$\Delta a^{\rm dir}_{CP}$ is not. The two-dimensional
significance is $2.0\sigma$, and thus the data is
inconsistent with \cp\ conservation at this level.

\Acknowledgements
We are grateful to the organizers of CHARM 2015 for an enjoyable
and productive workshop.


\begin{thebibliography}{99}
\bibitem{charge-conjugates}
Throughout this paper, charge-conjugate modes are implicitly
included unless stated or implied otherwise.
\end{thebibliography}
\end{document}